\documentclass[12pt]{article}
\usepackage{amsmath,epsf,amssymb,latexsym,cite,xy,shadow}
\xyoption{matrix}\xyoption{arrow}

\newtheorem{conj}{Conjecture}

\newif\iffigs\figstrue

\DeclareFontFamily{U}{rsf}{}
\DeclareFontShape{U}{rsf}{m}{n}{
  <5> <6> rsfs5 <7> <8> <9> rsfs7 <10-> rsfs10}{}
\DeclareMathAlphabet\Scr{U}{rsf}{m}{n}

\DeclareMathAlphabet\mathbi{U}{cmr}{bx}{it}

\def\pplogo{\vbox{\kern-\headheight\kern -29pt
\halign{##&##\hfil\cr&{%\sc
\ppnumber}\cr\rule{0pt}{2.5ex}&\ppdate\cr}
}}
\makeatletter
\def\ps@firstpage{\ps@empty \def\@oddhead{\hss\pplogo}%
  \let\@evenhead\@oddhead % in case an article starts on a left-hand page
}
\def\maketitle{\par
 \begingroup
 \def\thefootnote{\fnsymbol{footnote}}
 \def\@makefnmark{\hbox{$^{\@thefnmark}$\hss}}
 \if@twocolumn
 \twocolumn[\@maketitle]
 \else \newpage
 \global\@topnum\z@ \@maketitle \fi\thispagestyle{firstpage}\@thanks
 \endgroup
 \setcounter{footnote}{0}
 \let\maketitle\relax
 \let\@maketitle\relax
 \gdef\@thanks{}\gdef\@author{}\gdef\@title{}\let\thanks\relax}
\makeatother

\def\O{\Scr{O}}
\def\C{{\mathbb C}}
\def\P{{\mathbb P}}

\def\Z{{\mathbb Z}}

\def\Hom{\operatorname{Hom}}
\def\hom{\operatorname{hom}}

\def\Ext{\operatorname{Ext}}

\def\Sl{\operatorname{SL}}

\def\Cone{\operatorname{Cone}}

\def\ch{\operatorname{\mathit{ch}}}

\def\CY{Calabi--Yau}

\def\cA{{\Scr A}}

\def\cE{{\Scr E}}

\def\DC{\mathbf{D}}

\def\ff#1#2{{\textstyle\frac{#1}{#2}}}

\def\labto#1{\mathrel{\mathop\to^{#1}}}

\begin{document}
\setcounter{page}0
\def\ppnumber{\vbox{\baselineskip14pt
\hbox{DUKE-CGTP-02-07}
\hbox{hep-th/0209161}}}
\def\ppdate{September 2002} \date{}

\title{\LARGE Massless D-Branes on \CY\ Threefolds\\ and Monodromy\\[10mm]}
\author{
Paul S.~Aspinwall$^1$, R. Paul Horja$^2$, and Robert L.~Karp$^1$\\[4mm]
\normalsize $^1$Center for Geometry and Theoretical Physics \\
\normalsize Box 90318 \\
\normalsize Duke University \\
\normalsize Durham, NC 27708-0318 \\[8mm]
\normalsize $^2$Department of Mathematics\\
\normalsize University of Michigan\\
\normalsize East Hall, 525 E University Avenue\\
\normalsize Ann Arbor, MI 48109-1109\\[8mm]
}

{\hfuzz=10cm\maketitle}

\def\Large{\large}
\def\LARGE{\large\bf}

\vskip 1cm

\begin{abstract}
We analyze the link between the occurrence of massless B-type D-branes
for specific values of moduli and monodromy around such points in the
moduli space. This allows us to propose a classification of all
massless B-type D-branes at any point in the moduli space of \CY's.
This classification then justifies a previous conjecture due to Horja
for the general form of monodromy.  Our analysis is based on using
monodromies around points in moduli space where a single D-brane
becomes massless to generate monodromies around points where an
infinite number become massless. We discuss the various possibilities
within the classification.
\end{abstract}

\vfil\break

%%%%%%%%%%%%%%%%%%%%%%%%%%%%%%%%%%%%%%%%%%%%%%%%%%%%%%%%%%%%%%%%

\section{Introduction}    \label{s:intro}

The derived category approach to B-type D-Branes
\cite{Kon:mir,Doug:DC,Laz:DC,AL:DC,Dia:DC} appears to be extremely
powerful. It allows one to go beyond the picture of D-branes as vector
bundles over submanifolds so that $\alpha'$-corrections can be
correctly understood. For example, the fact that B-type D-branes must
undergo monodromy as one moves about the moduli space of complexified
K\"ahler forms can be expressed in the derived category
language \cite{Kont:mon,Horj:DX,Horj:EZ}.

The main purpose of this paper is to try to classify which D-branes
can become massless at a given point in the moduli space. Again the
language of derived categories will be invaluable.

In order for an object in the bounded derived category of coherent
sheaves to represent a D-brane it must be ``$\Pi$-stable''. Criteria
for $\Pi$-stability have been discussed in
\cite{DFR:stab,DFR:orbifold,Doug:Dgeom,AD:Dstab} although it is not
clear that we yet have a mathematically rigorous algorithm for
determining stability. Despite this, in simple examples such as in the
above references and \cite{me:point} one can compute stability with a
fair degree of confidence. In particular if you have reason to believe
that a certain set of a D-branes are stable at a given point in the
moduli space then one can move along a path in moduli space and see
how the spectrum of stable states changes. There is considerable
evidence \cite{AD:Dstab} that such changes in $\Pi$-stability depend
only on the homotopy class of the path in the moduli space of
conformal field theories.

The fact that changes in $\Pi$-stability {\em do\/} depend on the
homotopy class of such paths was used in \cite{AD:Dstab} to ``derive''
Kontsevich's picture of monodromy at least in the case of the quintic
\CY\ threefold.

The moduli space of conformal field theories may be compactified by
including the ``discriminant locus'' consisting of badly-behaved
worldsheet theories. Typically one expects such theories to be bad
because some D-brane has become massless \cite{Str:con}. Indeed, the
monodromy seen in \cite{Doug:DC,me:point} around parts of this
discriminant locus was intimately associated to massless D-branes.

It is this link between massless D-branes and monodromy that we wish
to study more deeply in this paper. In simple cases as one approaches
a point in the discriminant locus, a single D-brane becomes
massless. Of more interest to us is the case where an infinite number
become massless. 

In \cite{Horj:DX,Horj:EZ} one of the authors studied components of the
discriminant locus corresponding to what was called
``EZ-transformations''. Namely if one has a \CY\ threefold $X$ with
some complex subspace $E$, there may be a point in K\"ahler moduli
space where $E$ collapses to a complex subspace $Z$ of lower dimension
than $E$. We will see that it is then {\em the derived category of
$Z$\/} that describes the massless D-branes associated to this
transformation. A particular autoequivalence was naturally associated
to a particular EZ-transformation and it was conjectured in
\cite{Horj:DX,Horj:EZ} that such an autoequivalence resulted from the
associated monodromy. We will call this conjecture the ``EZ-monodromy
conjecture''. One purpose of this paper is to justify this conjecture.

Because of the nature of our understanding of D-branes and string
theory it will not be possible to rigorously {\em prove\/} any hard
theorems about D-branes. Instead we will have to play with a number of
conjectures whose interdependence leads to considerable evidence of
the validity of the overall story. In particular, on the one hand we
have the EZ-monodromy conjecture and, on the other hand,
we have our conjecture concerning which D-branes become
massless. These two conjectures are interlinked by $\Pi$-stability as
we discuss in section \ref{s:mon}. In particular, in section
\ref{ss:single} we discuss an older conjecture concerning single
massless D-branes. In section \ref{ss:gen} we then review a framework
for the more general case which is linked to the simpler case in
section \ref{ss:new} for a particular example. The physical
interpretation of the general case 
is then given in section \ref{ss:int}. 

The link discussed in section \ref{ss:new} between the simple case of
a single D-brane becoming massless and an infinite number becoming
massless depends upon a mathematical result which is derived in
section \ref{s:comp}. This section is more technical than the other
sections and may be omitted by the reader if need be. That said, it
shows how the sophisticated methods of derived categories are directly
relevant to the physics of D-branes.

In section \ref{s:app} we discuss a natural hierarchy of cases. The
familiar ``conifold''-like situation arises where $Z$ is a point and
only one soliton becomes massless. If $Z$ has dimension one then the
derived category of $Z$ has more structure. This case corresponds to
Seiberg--Witten theory of some nonabelian gauge group. We study an
explicit example of this elsewhere \cite{AK:Dsu2}.

The case where $Z$ has complex dimension two is more complicated as
the derived category now has a rich structure.  We show that it
appears to be similar to the spectrum of massless D-branes one gets
from a decompactification.  We also see that it demonstrates how
2-branes wrapped around a 2-torus can become massless. At first sight
this appears to contradict T-duality but we will see that this is not
actually the case.

Finally, for completeness, in section \ref{ss:exof} we discuss the
case of an exoflop which is awkward to fit into our general
classification but still yields a simple result.

%%%%%%%%%%%%%%%%%%%%%%%%%%%%%%%%%%%%%%%%%%%%%%%%%%%%%%%%%%%%%%%%%%%%

\section{Monodromy and Massless D-Branes} \label{s:mon}

\subsection{A Single Massless D-Brane} \label{ss:single}

B-type D-branes on $X$ correspond to objects in the bounded derived
category of coherent sheaves on $X$
\cite{Kon:mir,Doug:DC,Laz:DC,AL:DC,Dia:DC}. A given object
$\mathsf{A}$ is represented by a complex. We may then construct
another object $\mathsf{A}[n]$ by shifting this complex $n$ places to
the left. Such a shift or ``translation'' is a global symmetry of
physics if it is applied simultaneously to all objects
\cite{Doug:DC}. Relative shifts are significant --- an open string
stretched between $\mathsf{A}$ and $\mathsf{B}$ is not equivalent to an
open string stretched between $\mathsf{A}[n]$ and $\mathsf{B}$ if
$n\neq0$.

We would like to consider the case of  moving to a point in moduli
space where a single physical D-brane $\mathsf{A}$ becomes
massless. Because of the global shift symmetry all of its translates
$\mathsf{A}[n]$ are equally massless. Thus an infinite number of
objects in $\DC(X)$ are becoming massless even though only
one D-brane counts towards any physical effects of this masslessness
as it would be computed by Strominger \cite{Str:con} for example.

The analysis of $\Pi$-stability in \cite{AD:Dstab,me:point} showed
that monodromy is intimately associated to massless D-branes. This
should not be surprising since monodromy can only occur around the
discriminant and the discriminant is associated with singularities in the
conformal field theory associated with massless solitons
\cite{Str:con}.

Consider an oriented open string $f$ stretched between two D-branes in
a \CY\ threefold $X$. In
the derived category language this is written as a morphism between
two objects in $\DC(X)$
\begin{equation}
f:\mathsf{A}\to\mathsf{B}.
  \label{eq:fAB}
\end{equation}
These two objects may or may not form a bound state according to the
mass of the open string $f$. If $f$ is tachyonic then we have a bound
state \`a la Sen \cite{Sen:dbd}.
(As we emphasize shortly $\mathsf{A}$ is really an anti-brane in such
a bound state.)

A real number\footnote{It has been suggested that $\varphi$ is defined
modulo some integer such as 6 \cite{Doug:DC,DJP:dc6}. Periodicity can
also appear, if desired, in Floer cohomology 
(see \cite{Fuk:mir1,Sei:grd} for example) which
is supposedly mirror to the structure we are considering. For
simplicity we ignore such a possibility. To take such an effect into
account one should probably quotient the derived category by such
translations.}  (dubbed a ``grade'' in \cite{Doug:DC}) $\varphi$ is
associated to each stable D-brane. We assume $\varphi$ varies continuously
over the moduli space and is defined mod 2 by the central charge $Z$:
\begin{equation}
\varphi=-\frac1\pi\arg(Z)\pmod2.
\end{equation}
The precise definition of $\varphi$ is discussed at length in
\cite{AD:Dstab}.  In \cite{Doug:DC} it was argued that the mass
squared of the open string in (\ref{eq:fAB}) is then proportional to
$\varphi(\mathsf{B})-\varphi(\mathsf{A})-1$ allowing
the stability of this bound state to be determined.

One of the key features of the derived category which makes it so
useful for the study of solitons is the way that bound states are
described using {\em distinguished triangles}. The open string $f$
between $\mathsf{A}$ and $\mathsf{B}$ is best represented in the
context of a distinguished triangle
\begin{equation}
\xymatrix{
&\mathsf{C}\ar[dl]|{[1]}&\\
\mathsf{A}\ar[rr]^f&&\mathsf{B}.\ar[ul]
}   \label{eq:ABC}
\end{equation}
The ``$[1]$'' represents the fact that one must shift one place left
when performing the corresponding map.
The object $\mathsf{C}$, which is equivalent to the ``mapping cone''
$\Cone(f:\mathsf{A}\to\mathsf{B})$, is then potentially a bound state of
$\mathsf{A}[1]$ and $\mathsf{B}$. As explained in \cite{Doug:DC},
$\mathsf{A}[\mathrm{odd}]$ should be thought of as an
anti-$\mathsf{A}$.
The triangle also tells us that
$\mathsf{B}$ is potentially a bound state of $\mathsf{A}$ and
$\mathsf{C}$. Equally $\mathsf{A}$ is a bound state of $\mathsf{B}$
and $\mathsf{C}[-1]$. The ``$[1]$'' could be interpreted as keeping
track of which brane should be treated as an anti-brane.

The fact that $\DC(X)$ copes so well with anti-branes demonstrates its
power to analyze D-branes. The other approach, namely K-theory, should
be considered the derived category's weaker cousin since it only knows
about D-brane {\em charge\/}!

Now suppose that $\mathsf{A}$ is stable and becomes massless at
a particular point $P$ in the moduli space.  Furthermore, let us assume
that the {\em only\/} massless D-branes at $P$ are of the form $\mathsf{A}[m]$
for any $m$. Let us take a generic complex plane with polar
coordinates $(r,\theta)$ passing through $P$ at the origin and assume
that $Z(\mathsf{A})$ behaves as $cr\exp(-i\theta)$ near $P$ for $c$
some real and positive constant. That is, we assume that
$Z(\mathsf{A})$ has a simple zero at $P$.

Suppose $\mathsf{B}$ does not have vanishing mass. It follows that
$Z(\mathsf{B})$ and $Z(\mathsf{C})$ are equal at $P$ and nonzero. In
particular if we circle the point $P$ by varying $\theta$, these
central charges will be constant close to $P$. Furthermore, if
$\mathsf{C}$ can be a marginally bound state of anti-$\mathsf{A}$ and
$\mathsf{B}$ near $P$, then, according to the rules of \cite{AD:Dstab},
we have $\varphi(\mathsf{B})=\varphi(\mathsf{C})$ near $P$.

This allows us to rewrite (\ref{eq:ABC}) including the differences in
the $\varphi$'s for the open strings (i.e., sides of the triangle) to
give
\begin{equation}
\xymatrix{
&\mathsf{C}\ar[dl]_(0.4){1+a-b+\frac{\theta}{\pi}}|{[1]}&\\
\mathsf{A}\ar[rr]^f_{b-a-\frac{\theta}{\pi}}&&
  \mathsf{B},\ar[ul]_0
}   \label{eq:ABCg}
\end{equation}
where $\varphi(\mathsf{B})=b$ and $\varphi(\mathsf{A})=a$ at $\theta=0$.
The stability of a given vertex of this triangle depends upon the
number on the opposite side being less than 1. By ``stability'' we
mean relative to this triangle only. A given D-brane may decay by
other channels.

It follows that $\mathsf{C}$ becomes stable for $\theta>\pi(b-a-1)$
while $\mathsf{B}$ becomes unstable for $\theta>\pi(b-a)$.
Note that $\mathsf{A}$ is always stable near $P$ consistent with our
assumptions. 

Based on this idea that we ``gain'' $\mathsf{C}$ and ``lose''
$\mathsf{B}$ as $\theta$ increases, we can try to formulate a picture
for monodromy around $P$. The meaning of monodromy is that after
traversing this loop in the moduli space we should be able to relabel
the D-branes in such a way as to restore the physics we had before we
traversed the loop. It is important to note that monodromy is not
really the statement that a certain D-brane manifestly ``becomes''
another D-brane explicitly as we move through the moduli space. It is
much more accurately described as a relabeling process.

Since stability is a physical quality, we are forced to relabel
$\mathsf{B}$ since it has decayed. The obvious candidate in the above
case is to call it $\mathsf{C}$. Thus monodromy would transform
$\mathsf{B}$ into $\mathsf{C}$.

Life can be more complicated than this however. If we have an open
string $f':\mathsf{A}\to\mathsf{C}$, then, since
$\varphi(\mathsf{B})=\varphi(\mathsf{C})$ when $\mathsf{B}$ decays to 
$\mathsf{C}+\mathsf{A}[1]$, $\mathsf{C}$ will immediately decay further to
$\mathsf{D}=\Cone(f':\mathsf{A}\to\mathsf{C})$ plus another $\mathsf{A}[1]$.

Suppose $\mathsf{A}$ is ``spherical'' in the sense of \cite{ST:braid}
which means $\Hom(\mathsf{A},\mathsf{A}[m])=\C$ for $m=0$ or 3, and
$\Hom(\mathsf{A},\mathsf{A}[m])=0$ otherwise. This condition is always
satisfied in the context of this subsection --- i.e., only
$\mathsf{A}$ and its translates become massless. A long exact sequence
associated to (\ref{eq:ABC}) then implies
\begin{equation}
  \dim\Hom(\mathsf{A},\mathsf{C})=\dim\Hom(\mathsf{A},\mathsf{B})-1.
\end{equation}
It follows that this second decay will occur if
$\dim\Hom(\mathsf{A},\mathsf{B})>1$.  Iterating this process one sees
that $\mathsf{B}$ will decay splitting off an $\mathsf{A}[1]$ a total
of $\dim\Hom(\mathsf{A},\mathsf{B})$ times.

Finally we should also worry about homomorphisms between $\mathsf{B}$
and $\mathsf{A}[m]$ for other values of $m$. We refer to the example
in section 4 of \cite{me:point} for a detailed example of exactly how
this happens in a fairly nontrivial example.  All said, 
allowing for all these decays, $\mathsf{C}$ becomes a number of
$\mathsf{A}$'s (probably shifted) together with
\begin{equation}
  \Cone\left(\left(\ldots\oplus\mathsf{A}^{b_0}\oplus
        \mathsf{A}[-1]^{b_1}\oplus
        \mathsf{A}[-2]^{b_2}\oplus\ldots\right)\to\mathsf{B}\right),
	\label{eq:Aaz}
\end{equation}
where 
\begin{equation}
\begin{split}
  b_n &= \dim\Hom(\mathsf{A}[-n],\mathsf{B})\\
      &= \dim\Ext^n(\mathsf{A},\mathsf{B}).
\end{split}
\end{equation}
The cone (\ref{eq:Aaz}) may be written more compactly as
\begin{equation}
  \mathbi{K}_{\mathsf{A}}(\mathsf{B}) =
  \Cone(\hom(\mathsf{A},\mathsf{B})\otimes\mathsf{A}\to\mathsf{B}),
	\label{eq:cone1}
\end{equation}
where $\hom(\mathsf{A},\mathsf{B})$ is the complex of $\C$-vector
spaces
\begin{equation}
  \ldots \labto0 \Ext^0(\mathsf{A},\mathsf{B}) \labto0 
  \Ext^1(\mathsf{A},\mathsf{B}) \labto0 
  \Ext^2(\mathsf{A},\mathsf{B}) \labto0 \ldots
\end{equation}
We refer to \cite{ST:braid} for further explanation of the
notation.\footnote{Note that since ``left-derived'' {\bf L}'s or
``right-derived'' {\bf R}'s should be added to every functor in this
paper, we may consistently omit them without introducing any ambiguities!} 
We can also write more heuristically
\begin{equation}
  \renewcommand{\sboxsep}{6pt}
  \mathbi{K}_{\mathsf{A}}(\mathsf{B}) = \Cone\left(
  \shabox{\parbox{37mm}{\scriptsize As much massless stuff that can bind
      to $\mathsf{B}$ as possible.}}
  \to\mathsf{B}\right).
  \label{eq:heur}
\end{equation}

Interpreted na\"\i vely, we have shown that, upon increasing $\theta$
from $-\infty$ to $+\infty$, an object $\mathsf{B}$ will decay and 
a canonically associated object $\mathbi{K}_{\mathsf{A}}(\mathsf{B})$
will become stable and appear as one of the decay products of
$\mathsf{B}$. What is desired however is monodromy {\em once\/} around
$P$, i.e., $\theta$ should only increase by $2\pi$. We will indeed
claim that monodromy once around $P$ replaces $\mathsf{B}$ by
$\mathbi{K}_{\mathsf{A}}(\mathsf{B})$.

For some objects, increasing $\theta$ by only
$2\pi$ (at the appropriate starting point) will cause the complete
decay of $\mathsf{B}$ into
$\mathbi{K}_{\mathsf{A}}(\mathsf{B})$. Thanks to its rather simple
cohomology, this always happens for $\O_x$, the structure sheaf of a
point $x\in X$.
Therefore the relabeling process under monodromy should replace
$\mathsf{B}$ by $\mathbi{K}_{\mathsf{A}}(\mathsf{B})$.
There are undoubtedly many other objects $\mathsf{B}'$ under which
this increase in $\theta$ by only $2\pi$ would not induce the entire
decay to $\mathbi{K}_{\mathsf{A}}(\mathsf{B}')$. This doesn't matter
however, we can still leave physics invariant by relabeling
$\mathsf{B}'$ by $\mathbi{K}_{\mathsf{A}}(\mathsf{B}')$. For example
in an extreme case, both $\mathsf{B}'$ and
$\mathbi{K}_{\mathsf{A}}(\mathsf{B}')$ may be stable with respect to
the above triangle both before and after increasing $\theta$ by
$2\pi$. It is therefore harmless to relabel one of these states as the
other.

Well, it is fine saying that it is harmless to relabel $\mathsf{B}'$ by
$\mathbi{K}_{\mathsf{A}}(\mathsf{B}')$, but why are we {\em forced\/}
to relabel like this? The reason is that we know that physics must be
completely invariant under monodromy which implies that the relabeling
must amount to an {\em autoequivalence\/} of $\DC(X)$.
One can indeed show that $\mathbi{K}_{\mathsf{A}}$
defines an autoequivalence\footnote{Pedants will object that the cone
construction is only defined up to a non-canonical isomorphism making
the transformation on morphisms badly-defined. Fortunately, as is
well-known and we discuss at length in section \ref{s:comp}, this
transformation can be written as a Fourier--Mukai transform removing
this objection.} of $\DC(X)$ so long as $\mathsf{A}$ is spherical
\cite{ST:braid,Horj:EZ}. What's more, as argued in
\cite{AD:Dstab,BM:FMq}, it is pretty well the {\em only\/}
autoequivalence that works. To be more precise, once we have argued
that the specific objects $\O_x$ undergo monodromy given by
$\mathbi{K}_{\mathsf{A}}$ then all other objects must undergo the same
monodromy up to some possible multiplication by some fixed line bundle $L$.

Note that the central charge $Z$ is also a physical
quantity. Insisting that monodromy acts correctly in this case amounts
to insisting that the D-brane ``charges'' $\ch(\mathsf{B})$ transform
under monodromy.  This is precisely the same monodromy on
$H^{\mathrm{even}}(X,\Z)$ that one deduces from mirror symmetry as in
\cite{CDGP:}.  This determines that the above line bundle $L$ is
trivial (in a considerably overdetermined way!).  It is known (see
\cite{Horj:DX} for example) that $\mathbi{K}_{\mathsf{A}}$ then induces the
correct transformation on these charges --- indeed this was the reason why
$\mathbi{K}_{\mathsf{A}}$ was conjectured as the monodromy action in
the first place \cite{Kont:mon}!

It is worth noting that in some special cases the transformation
$\mathbi{K}_{\mathsf{A}}$ has nothing to do with decay. Consider how
the spherical object $\mathsf{A}$ itself transforms:
\begin{equation}
\begin{split}
  \Cone(\hom(\mathsf{A},\mathsf{A})\otimes\mathsf{A}\to\mathsf{A}) &=
    \Cone((\underline{\C}\to0\to0\to\C)\otimes\mathsf{A}\to\mathsf{A})\\
  &= \Cone((\underline{\mathsf{A}}\to0\to0\to\mathsf{A})\to\mathsf{A})\\
  &\cong \underline{0}\to0\to\mathsf{A}\\
  &= \mathsf{A}[-2],
\end{split}   \label{eq:Ocon}
\end{equation}
where we use the convention of \cite{AL:DC} by underlining the zero
position when necessary. Such a transformation cannot be argued from
$\Pi$-stability however. Clearly an open string between $\mathsf{A}$
and itself (perhaps translated) cannot have a mass that depends upon
some angle as we orbit the conifold point as clearly the mass is
constant. Instead one could argue that the transform (\ref{eq:Ocon})
occurs simply because $Z(\mathsf{A})$ has a simple zero at the
conifold point and thus $\varphi(\mathsf{A})$ shifts by $-2$ as we
loop around the conifold point. Then we can apply the rule
$\varphi(\mathsf{A}[n])=\varphi(\mathsf{A})+n$ from
\cite{AD:Dstab}.\footnote{In \cite{DJP:dc6} it was suggested that the
monodromy action on the derived category should be translated by 2 to
undo this action on $\mathsf{A}$. Since monodromy is a relabeling
process, one is free to do this, but it looks unnatural from the
perspective of associating monodromy with $\Pi$-stability.}

We must therefore view (\ref{eq:cone1}) as being motivated by
$\Pi$-stability for {\em most\/} but not all of the objects in
$\DC(X)$. Note that the fact that the obvious physical requirement
that monodromy be an autoequivalence of $\DC(X)$ can force
(\ref{eq:cone1}) to be the required transform for {\em all\/} the
objects in $\DC(X)$ once $\Pi$-stability has established it for a few
elements. This was the basis of the proof in the case of the quintic
in \cite{AD:Dstab}.

Anyway, all said we have motivated the following conjecture (which, in
perhaps a slightly different form, is due to Kontsevich
\cite{Kont:mon}, Horja \cite{Horj:DX} and Morrison \cite{Mor:geom2}):
\begin{conj}
If we loop around a component of the discriminant locus associated
with a single D-brane $\mathsf{A}$ (and thus its translates) becoming
massless then this results in a relabeling of D-branes given by an
autoequivalence of the derived category in which $\mathsf{B}$ becomes
$\Cone(\hom(\mathsf{A},\mathsf{B})\otimes\mathsf{A}\to\mathsf{B})$.
\label{conj:a}
\end{conj}
This transformation was also motivated by its
relation to mirror symmetry and studied at length by Seidel and Thomas
\cite{ST:braid}. 

%%%%%%%%%%%%%%%
\subsection{General Monodromies} \label{ss:gen}

It is then natural to ask what happens more generally, i.e., if more
than just a single D-brane $\mathsf{A}$ becomes massless. In order to answer this we
need to set up a general description of how one might analyze
monodromy in a multi-dimensional moduli space.

There are two paradigms for monodromy --- both of which are useful:
\begin{enumerate}
\item The discriminant locus decomposes into a sum of irreducible
  divisors. Pick some base point in the moduli space and loop around a
  component of the discriminant ``close'' to the base point.
\item Restrict attention to a special rational curve $C$ in the moduli
  space. This rational curve contains two ``phase limit points'', in a sense to
  be described below, and a single point in the discriminant. The
  loop in question is around this unique discriminant point.
\end{enumerate}
In the case of the one parameter models, such as the quintic, these
two paradigms coincide. The moduli space {\em is\/} $C\cong\P^1$ and the
discriminant locus is a single point.
If a component of the discriminant intersects $C$ transversely then we
can again have agreement between these two pictures of monodromy. In
general the discriminant need not intersect $C$ transversely --- a
fact we use to our advantage in section \ref{ss:new}.

We now recall the relationship between the discriminant locus and
phases as analyzed in \cite{AGM:sd,MP:inst}. The following is
a very rapid review. Please refer to the references for more details.

To make the discussion easier we suffer a little loss of generality
and assume we are in the ``Batyrev-like'' \cite{Bat:m} case $X$ being
a hypersurface in a toric variety. The data for $X$ is then presented
in the form of a point set $\cA$ which is the intersection of some
convex polytope with some lattice $N$. See \cite{AG:gmi}, for example,
for more details of this standard construction. The conformal field
theory associated to this data then has a phase structure where each
``phase'' is associated to a regular triangulation of $\cA$
\cite{W:phase,AGM:I}. The real vector space in which the K\"ahler form
lives is naturally divided into a ``secondary fan'' of all possible
phases. One cone of this fan is the K\"ahler cone for $X$ where we
have the ``\CY'' phase.

Mirror to $X$, $Y$ is described as the zero-set of a polynomial $W$ in
many variables. The points in $\cA$ are associated one-to-one with
each monomial in $W$. Thus the data $\cA$ is associated to
deformations of complex structure of $Y$ via the monomial-divisor
mirror map \cite{AGM:mdmm}.

If we model the moduli space of complex structures on $Y$ by the space
of coefficients in $W$, then the discriminant locus can be computed by the
failure of $W$ to be transversal. This can be mapped back to the space of
complexified K\"ahler forms on $X$. The result is that part of the
discriminant asymptotically lives in each wall dividing adjacent
phases in the space of K\"ahler forms. That is to say, if we tune the
$B$-field suitably we can always hit a bad conformal field theory as
we pass from one phase to another. Thus we may associate singular
conformal field theories with phase transitions.

The discriminant itself is generically reducible. The combinatorial
structure of this reduction has been studied in detail in
\cite{GKZ:book}. In particular, any time an $m$-dimensional face of the
convex hull of the set $\cA$ contains more than $m+1$ points, the
resulting linear relationship between these points yields a component
of $\Delta$. One may then follow an algorithm presented in
\cite{MP:inst} to compute the explicit form of each component.
The general picture then is of a discriminant with many
components with each component having ``fingers'' which separate the
phases from each other. Each phase transition is associated with
fingers from one or more component of $\Delta$. 

Torically each maximal cone in the secondary fan is associated to a
point in the moduli space which gives the limit point in the ``deep
interior'' of the associated phase. The real codimension-one wall
between two maximal cones corresponds to a rational curve $C$ passing
through two such limit points. The rational curve $C$ will intersect
the discriminant locus in one point as promised earlier in this section.

One component of $\Delta$ is distinguished --- it corresponds to the
case of viewing the full convex hull as a face of itself. This is
called the ``primary'' component of $\Delta$. Closely tied in with
conjecture~\ref{conj:a} (and at least partially attributed to the
same authors) is the following conjecture 
\begin{conj}
At any point on the primary component of $\Delta$ (reached by a
suitable path from a suitable basepoint) the 6-brane
associated with the structure sheaf $\O_X$ and its translates become
massless. At a generic point no other D-branes become massless.
\label{conj:b}
\end{conj}
This idea was perhaps first discussed in \cite{GK:vol}. It is
certainly a very natural conjecture --- the primary component of the
discriminant is a universal feature for any \CY\ manifold and so must
be associated with the masslessness of a very basic D-brane. The fact 
that it
works for the quintic was explicitly computed in \cite{AD:Dstab}, and
presumably it is possible to verify the conjecture in a much larger
class of examples. We will assume this conjecture to be true.

The K\"ahler cone is a particular maximal cone in the secondary fan
corresponding to the ``\CY'' phase.  Let us concentrate on the walls
of the K\"ahler cone. A typical situation as we approach the wall of
the K\"ahler cone is that an exceptional set $E$ collapses to some
space $Z$. We depict this as
\begin{equation}
\xymatrix{
E\, \ar[d]^q\ar@{^{(}->}[r]^i &X\\
Z
} \label{eq:EZ}
\end{equation}
where $i$ is an inclusion (which may well be the identity) and $q$ is
a fibration with a strict inequality $\dim(E)>\dim(Z)$. 

Associated with such a wall in the secondary fan we have a rational
curve $C$ in the moduli space connecting the large radius limit point
with some other limit point. We wish to consider the monodromy
associated to circling the point in the discriminant in $C$.  The
resulting autoequivalence on the derived category 
has been studied in \cite{Horj:EZ} where it was dubbed an
``EZ-transformation''.

The simplest example would be the case of the quintic \CY\
threefold which has only one deformation of the K\"ahler form. This
single component of the K\"ahler form gives the overall size of the
manifold. Thus the ``wall'' (i.e., the origin) corresponds to $X$
collapsing to a point. In this case $i$ is the identity map and $Z$ is
a point. This is the case discussed above in section \ref{ss:single}.

Indeed, it appears that in all cases where $Z$ is a point, the
resulting monodromy amounts to a transform of the type studied in
section \ref{ss:single}. It is precisely when $Z$ is more than just a
point the case of interest to us.

%%%%%%%%%%%

\subsection{New monodromies from old} \label{ss:new}

The precise form of the ``EZ-monodromy conjecture'' which associates
an autoequivalence of $\DC(X)$ with a given EZ-transform was given in
\cite{Horj:EZ}. Rather than appealing this conjecture, let us derive
the simplest example of a more general case, by assuming the
conjectures above, dealing with the case of a single massless D-brane.

We will look at the well-known example \cite{CDFKM:I}, where $X$ is a
degree 8 hypersurface in the resolution of a weighted projective space
$\P^4_{\{2,2,2,1,1\}}$. The mirror $Y$ is then a quotient of the same
hypersurface with defining equation
\begin{equation}
  a_0z_1z_2z_3z_4z_5 + a_1z_1^4 + a_2z_2^4 + a_3z_3^4 + a_4z_4^8
  + a_5z_5^8 + a_6z_4^4z_4^4.  \label{eq:def1}
\end{equation}
The ``algebraic'' coordinates on the moduli space are then given by
\begin{equation}
  x=\frac{a_1a_2a_3a_6}{a_0^4},\quad y=\frac{a_4a_5}{a_6^2}.
        \label{eq:algco}
\end{equation}
The primary component of $\Delta$ can  be computed as
\begin{equation}
  \Delta_0 = (1-2^8x)^2 - 2^{18}x^2y.
\end{equation}
The edge of the convex hull containing the points labeled by
$a_4,a_5,a_6$ leads to another component
\begin{equation}
  \Delta_1 = 1-4y,
\end{equation}
with $\Delta=\Delta_0\Delta_1$.

$X$ can be viewed as a K3-fibration $\pi:X\to\P^1$. In this case the
component of the 
K\"ahler form given asymptotically (for $x,y\ll1$) by $\frac1{2\pi
i}\log(x)$ controls the size of the K3 fibre. The component of the
K\"ahler form given asymptotically by $\frac1{2\pi i}\log(y)$ gives
the size of the $\P^1$ base.

\iffigs
\begin{figure}
  \centerline{\epsfxsize=9cm\epsfbox{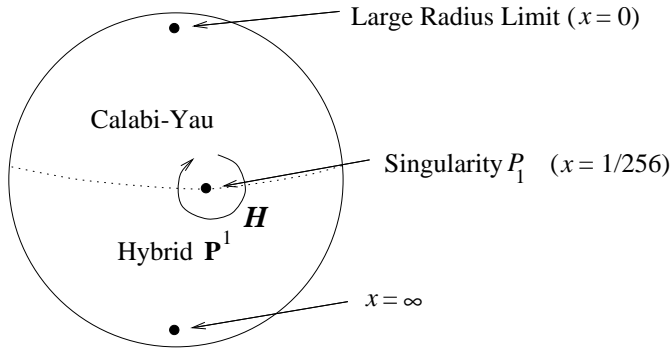}}
  \caption{Moduli Space for $y=0$.}
  \label{f:y0}
\end{figure}
\fi

The base $\P^1$ is made very large by setting $y\to0$. In
this case, we hit the primary component $\Delta_0$ of the discriminant
when $x=2^{-8}$. Let us refer to this point as $P_1$.
Increasing $x$ beyond this value moves one out of the
\CY\ phase into the
hybrid ``$\P^1$-phase'' where the model is best viewed as a fibration
with base $\P^1$ and a Landau--Ginzburg orbifold as fibre
\cite{AGM:I}. Fixing $y=0$ and varying $x$ spans a rational curve $C$ in the
moduli space shown in figure~\ref{f:y0}.

We would like to analyze the monodromy around the singularity $P_1$ in
figure~\ref{f:y0}. Clearly the transition associated with this
monodromy consists of collapsing $X$ onto the $\P^1$ base. In the
language of (\ref{eq:EZ}), $E=X$, i.e., the inclusion map $i$ is the
identity, and $Z\cong\P^1$. The map $q$ is given by the fibration map $\pi$.
In a way, we have constructed the simplest possible example where $Z$ is more
than just a point.

Now the useful trick is that the monodromy around the singularity in
figure~\ref{f:y0} can be written in terms of other monodromies that we
already understand. This was originally described in
\cite{CDFKM:I}, while this feature was also exploited in
\cite{Horj:DX,me:navi}. 

The full moduli space near $P_1$ is shown in figure~\ref{f:full} (with
complex dimensions shown as real). The rational curve $C$ given by
$y=0$ corresponds to an infinite radius limit and as such we
understand the monodromy around it (see for example
\cite{Horj:DX,me:navi}). (We will follow closely the notation and
analysis of \cite{me:navi}).

Let $\mathbi{L}$ refer to the autoequivalence of
$\DC(X)$ we apply upon looping this curve. It follows that
\begin{equation}\label{eq:defl}
  \mathbi{L}(\mathsf{B}) = \mathsf{B}\otimes\O_X(S),
\end{equation}
where $S$ is the divisor class of a K3 fibre in $X$.
Meanwhile let $\mathbi{K}$ refer to the autoequivalence of
$\DC(X)$ we apply upon looping the primary component
$\Delta_0=0$ (i.e., denote $\mathbi{K}_{\O_X}$ of section
\ref{ss:single} by $\mathbi{K}$).
Then from conjectures \ref{conj:a} and \ref{conj:b} we know
that
\begin{equation}
  \mathbi{K}(\mathsf{B}) = \Cone(\hom(\O_X,\mathsf{B})\otimes\O_X
     \to\mathsf{B}). \label{eq:mO}
\end{equation}

It follows (see, for example section 5.1 of \cite{me:navi} for an
essentially identical computation) that the autoequivalence
%$\mathbi{H}$,
for the desired loop shown in figure~\ref{f:y0} around $P_1$ is given
by
\begin{equation} \label{eq:defh}
%  \mathbi{H} = 
\mathbi{L}^{-1}\mathbi{K}\mathbi{L}\mathbi{K}.
\end{equation}
The desired goal therefore is to find the autoequivalence of $\DC(X)$
obtained by combining the transforms in the above form.

\iffigs
\begin{figure}
  \centerline{\epsfxsize=12cm\epsfbox{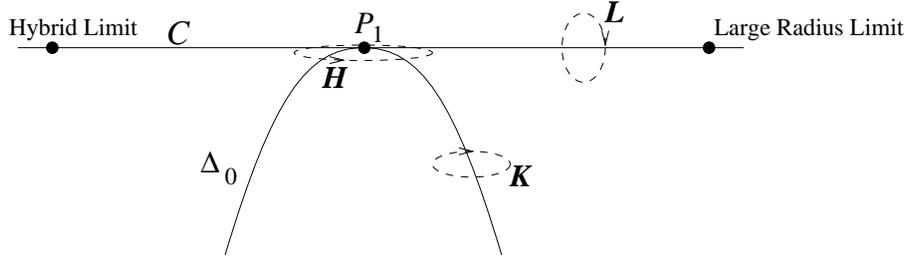}}
  \caption{Full Moduli Space around $P_1$.}
  \label{f:full}
\end{figure}
\fi

The result is that 
\begin{equation} 
\mathbi{L}^{-1}\mathbi{K}\mathbi{L}\mathbi{K}=\mathbi{H},
\label{eq:pf}
\end{equation}
where $\mathbi H$ is an autoequivalence that 
acts on $\DC(X)$ by 
\begin{equation}
  \mathbi{H}(\mathsf{B}) = \Cone(\pi^*\pi_*\mathsf{B}\to\mathsf{B}).
	\label{eq:picon}
\end{equation}
Section \ref{s:comp} is devoted to the proof of this statement.
Let us review briefly  what is exactly meant by
the rather concise notation of (\ref{eq:picon}). Given the map
$\pi:X\to Z$ and a sheaf $\cE$ on $X$ we may construct the
``push-forward'' sheaf $\pi_*\cE$ on $Z$ by associating $\pi_*\cE(U)$
with $\cE(\pi^{-1}U)$ for any open set $U\subset Z$. The $\pi_*$
appearing in (\ref{eq:picon}) is the right-derived functor of this
push-forward map. This $\pi_*$ ``knows'' about the cohomology of the
fibre of $\pi$ (see, for example, chapter III of \cite{Hartshorne:}).
The pull-back map $\pi^*$ is defined for sheaves of $\O_Z$-modules,
and in particular for locally free sheaves, and thus for
vector bundles. The map $\pi^*$ appearing in (\ref{eq:picon}) is
the corresponding left-derived functor. It is a central result of the
theory of derived categories \cite{Hart:dC} that $\pi^*$ is the
left-adjoint of $\pi_*$:
\begin{equation}
  \Hom_X(\pi^*\mathsf{E},\mathsf{F}) \cong
  \Hom_Z(\mathsf{E},\pi_*\mathsf{F}),
  	\label{eq:ladj}
\end{equation}
for any $\mathsf{E}\in\DC(Z)$ and $\mathsf{F}\in\DC(X)$. It follows
that 
\begin{equation}
  \Hom_X(\pi^*\pi_*\mathsf{B},\mathsf{B}) \cong
  \Hom_Z(\pi_*\mathsf{B},\pi_*\mathsf{B}).
	\label{eq:id1}
\end{equation}
Thus the most natural morphism that would appear in (\ref{eq:picon})
is the image of the identity on the right-hand side of equation
(\ref{eq:id1}) under this natural isomorphism. One can show that this
is indeed the case.

%%%%%%%%%%%%%%%%%%%%%%%%%%%%%%%

\subsection{Interpretation of monodromy} \label{ss:int}

Let us interpret (\ref{eq:picon}) in light of our discussion of
monodromy from $\Pi$-stability in section \ref{ss:single}. To aid our
discussion consider how one might rewrite the monodromy result
(\ref{eq:mO}) for the primary component of the discriminant. Let 
$c:X\to x$ be the constant map of $X$ to a single point. One can then
show, using the fact that sheaf cohomology is equivalent to $c_*$,
which, in turn, is also given by the global section functor
$\Hom(\O_X,-)$ \cite{Hartshorne:}, that (\ref{eq:mO}) is equivalent to
\begin{equation}
  \mathbi{K}(\mathsf{B}) = \Cone(c^*c_*\mathsf{B} \to \mathsf{B}).
	\label{eq:mOc}
\end{equation}
Now, the only D-brane (up to translation in $\DC(X)$) which becomes
massless in this case is $\O_X$, which is equal to $c^*\C$, where we
denote the trivial (very trivial!) line bundle on the point $x$ as
$\C$. That is, massless D-branes for the primary component of the
discriminant are given by $c^*$(something). The $c_*$ in
(\ref{eq:mOc}) then gives a natural map to form a cone as required.

The expression~(\ref{eq:heur}) immediately dictates that we
may interpret $\Cone(\pi^*\pi_*\mathsf{B}\to\mathsf{B})$ in a
similar way. {\it The D-branes becoming massless at the point $P_1$ in
the moduli space correspond to $\pi^*\mathsf{z}$ for some
$\mathsf{z}\in \DC(Z)$.} The push-forward
map $\pi_*$ can be viewed as the natural ingredient required to form
the cone from (\ref{eq:id1}).

There is one technical subtlety here which needs to be mentioned. The
set of objects of the form $\pi^*\mathsf{z}$ for any
$\mathsf{z}\in\DC(Z)$ is not closed under composition by the cone
construction. If we write
\begin{equation}
  \mathsf{C} = \Cone(f:\pi^*\mathsf{a}\to\pi^*\mathsf{b}),
\end{equation}
for two objects $\mathsf{a},\mathsf{b}\in\DC(Z)$, then we may only
write
\begin{equation}
  \mathsf{C} = \pi^*\Cone(f':\mathsf{a}\to\mathsf{b}),
\end{equation}
when there is a relationship between the morphisms
$f=\pi^*f'$. Unfortunately such an $f'$ need not exist for arbitrary
$f$. Thus we might more properly say that the set of massless D-branes
are {\em generated\/} by objects of the form $\pi^*\mathsf{z}$ where
we allow for composing such states.

Flushed with success at interpreting the autoequivalence given by this
example we now write down the most obvious generalization for the more
general EZ-transform (\ref{eq:EZ}). First of all, we expect
$q^*$(something) to be a massless D-brane on $E$. $E$ is mapped into
$X$ by the inclusion map $i$. The push-forward map $i_*$ is then
``extension by zero'' of a sheaf which is the obvious way of mapping
D-branes on $E$ into D-branes on $X$. Our next conjecture is then
\begin{conj}
Any D-Brane which becomes massless at a point on a component of the
discriminant associated with an EZ-transform is generated by objects
of the form $i_*q^*\mathsf{z}$ for $\mathsf{z}\in\DC(Z)$.
  \label{c:massless}
\end{conj}

This implies a corresponding autoequivalence for the monodromy from
$\Pi$-stability:
\begin{equation}
  \mathsf{B} \mapsto \Cone(i_*q^*\zeta\mathsf{B}\to\mathsf{B}),
\end{equation}
for some ``natural'' map $\zeta:\DC(X)\to\DC(Z)$. To compute $\zeta$ we
use the same trick as (\ref{eq:id1}). Introduce the functor ``$i^!$'' as the
right-adjoint of $i_*$:
\begin{equation}
  \Hom_X(i_*\mathsf{E},\mathsf{F}) \cong
  \Hom_E(\mathsf{E},i^!\mathsf{F}),
  	\label{eq:radj}
\end{equation}
for any $\mathsf{E}\in\DC(E)$ and $\mathsf{F}\in\DC(X)$. 
The existence of $i^!$ is one of the most important features of the
derived category in algebraic geometry \cite{Hart:dC} and
(\ref{eq:radj}) may be regarded as a generalization of Serre Duality. 
Now we have 
\begin{equation}
  \Hom_X(i_*i^!\mathsf{B},\mathsf{B}) \cong
  \Hom_E(i^!\mathsf{B},i^!\mathsf{B}).
	\label{eq:id2}
\end{equation}
This leads to a natural map $i_*q^*q_*i^!\mathsf{B}\to
i_*i^!\mathsf{B}\to\mathsf{B}$. This implies
\begin{conj}
The monodromy around the discriminant in the wall associated to a
phase transition given by an EZ-transform leads to the following
autoequivalence on $\DC(X)$
\begin{equation}
  \mathsf{B} \mapsto \Cone(i_*q^*q_*i^!\mathsf{B}\to\mathsf{B}).
	\label{eq:EZm}
\end{equation}
  \label{c:EZ}
\end{conj}
This is equivalent to the EZ-monodromy conjecture of
\cite{Horj:EZ}. In particular, it was proven there that this is indeed
an autoequivalence of $\DC(X)$.
We should perhaps emphasize that we have not {\em
proven\/} conjectures \ref{c:massless} and \ref{c:EZ}. Rather we have
used the known connection between $\Pi$-stability and monodromy and
generalized the example we considered in the simplest and most obvious
way. 

Note that we may derive conjecture \ref{conj:a} from conjectures
\ref{c:massless} and \ref{c:EZ} as follows. Suppose $Z$ is a point,
then $i_*q^*\mathsf{z}$ can only be one thing, so we have a single
massless D-brane. Furthermore, $q_*$ now becomes the cohomology
functor giving
\begin{equation}
\begin{split}
  i_*q^*q_*i^!\mathsf{B} &= i_*(\hom_E(\O_E,i^!\mathsf{B})\otimes\O_E)\\
    &= \hom_X(i_*\O_E,\mathsf{B})\otimes i_*\O_E.
\end{split}
\end{equation}
This also shows that the massless D-brane is $i_*\O_E$ --- i.e., the
D-brane wrapping around $E$ as observed in
\cite{DFR:orbifold,DG:fracM,Doug:DC}.

For completeness we should also obtain the monodromy as one circles
the discriminant in the opposite direction. 
Going back to the triangle (\ref{eq:ABCg}) we see that {\em
decreasing\/} $\theta$ would result in $\mathsf{C}$ being replaced
by $\mathsf{B}=\Cone(\mathsf{C}\to\mathsf{A}[1])[-1]$.
This implies we modify the above monodromy arguments to consider the
transformation under which
\begin{equation}
  \mathsf{C} \mapsto \Cone(\mathsf{C}\to i_*q^*\eta\mathsf{C})[-1],
\end{equation}
for some ``natural'' map $\eta:\DC(X)\to\DC(Z)$. That is, the massless
objects bind ``to the right'' of $\mathsf{C}$ in the mapping cone rather than
to the left. The ``$[-1]$'' is needed because of the asymmetrical
definition of the mapping cone --- we need to keep $\mathsf{C}$ in its
original position.

The only nontrivial step in copying the above argument is that to
construct $\eta$ we need a left-adjoint functor for $q^*$. Given that
$q^!\mathsf{F} = q^*\mathsf{F}\otimes q^!\O_Z$ we can construct such a
functor from
\begin{equation}
\begin{split}
  \Hom_E(\mathsf{E},q^*\mathsf{F}) &= \Hom_E(\mathsf{E}\otimes
  q^!\O_Z,q^!\mathsf{F}) \\
  &= \Hom_Z(q_*(\mathsf{E}\otimes q^!\O_Z),\mathsf{F}).
\end{split}
\end{equation}
Thus our desired transform is given by
\begin{equation}
  \mathsf{C} \mapsto \Cone\left(\mathsf{C}\to i_*q^*q_*
  (i^*\mathsf{C}\otimes q^!\O_Z)\right)[-1].
\end{equation}
It was shown in \cite{Horj:EZ} that this is indeed the inverse of the
transformation given in conjecture~\ref{c:EZ}.

%%%%%%%%%%%%%%%%%%%%%%%%%%%%%%%%%%%%%%%%%%%%%%%%%%%%%%%%%%%%%%%%%%

\section{Composing transforms} \label{s:comp}

In this section we will prove (\ref{eq:pf}). For completeness, we 
choose to adopt a more general point of view and work with the class
of Calabi--Yau fibrations over projective spaces. Thus we cover
examples such as elliptic fibrations over $\P^2$ as discussed in
section \ref{eq:sf}, as well as the case of K3 fibrations over $\P^1$
as desired in section \ref{ss:new}.

Readers not familiar with manipulations in the derived category may
well wish to accept the result and skip this section. Having said
that, some of the methods used in this section are very powerful and
may have many other applications to D-brane physics.

For the sake of brevity we use the formalism of kernels to describe
the Fourier-Mukai transforms. For the convenience of the reader we
review some of the key notions involved. The notations follow those of
\cite{Horj:EZ}.

For $X$ a non-singular projective variety, an object ${\cal
  G}\in\DC(X\times X)$ determines an exact functor of triangulated categories 
$\Phi_{\cal G} : \DC(X) \to\DC(X) $ by the formula
\begin{equation} \label{eq:ker}
\Phi_{\cal G}(-):= p_{2_*}
({\cal G}\otimes  p_1^*(-)),
\end{equation}
where $p_1 : X \times X \to X$ is projection on the first factor,
while $p_2$ is projection on the second factor.
The object ${\cal G}\in\DC(X\times X)$ is called the {\it kernel}. 

The convenience in using kernels comes about because of the following
natural isomorphism of functors
\begin{equation}\label{f1}
\Phi_{{\cal G}' \star \,{\cal G}} \cong \Phi_{{\cal G}'} \circ \Phi_{{\cal G}}.
\end{equation}
The composition of the kernels ${\cal  G}',{\cal G} \in \DC(X\times X)$ 
is defined as 
\begin{equation} \label{def:comp}
{\cal G}'\star \,{\cal G }  := p_{13_*} \big( p_{23}^* ( {\cal G}')\otimes
p_{12}^* ({\cal G}) \big),
\end{equation}
where $p_{ij}$ is the obvious projection from $X\times X\times X$ to
the relevant two factors.
There is an identity element for the composition of kernels:
$(\Delta_X)_* (\O_X),$ where $\Delta_X : X \hookrightarrow X \times X$
is the diagonal morphism.

In this section $X$ is assumed to be a smooth Calabi--Yau fibration 
%(projective variety with trivial canonical bundle)
of dimension $n$ over $Z \cong \P^d,$ with 
$\pi:X\to Z$ the fibration map.
For us, the Calabi--Yau fibration structure simply 
means that $\pi:X\to Z$ 
is a flat morphism (see section III.9 of
\cite{Hartshorne:}) with the generic fibre 
a Calabi--Yau variety of dimension $n-d.$ Further 
assumptions on the Calabi--Yau fibration will be added shortly.

In order to set the functors $\mathbi{L},$ $\mathbi{K}$
and $\mathbi{H}$ of the previous section on firm mathematical 
footing and to define them in the more general context of this section,  
we need to describe
the kernels that induce them as exact functors according to
formula (\ref{eq:ker}). 
The following commutative diagram contains most of the maps that we use in
the sequel:
\begin{equation}\label{c1}
\xymatrix@=15mm{
  &  *++{X}   \ar@{^{(}->}[d]^j           & \\
X\ar[d]_\pi & X \times X \ar[l]_{p_1} \ar[r]^{p_2} \ar[ld]_{\pi_1}
  \ar[dr]^{\pi_2}\ar[d]^{\pi\times\pi} &X\ar[d]^\pi \\
Z & Z\times Z\ar[l]^{s_1} \ar[r]_{s_2}&  Z\,.
}
\end{equation}  
The maps are mostly projections, that are obvious from the context,
except for $j:=\Delta_X : X \hookrightarrow X \times X$ the
diagonal of $X$ and $\pi:X\to Z$ the fibration map.
%In this section, we sometimes choose to write $\O$ without a subscript
%for $\O_Z$, and similarly for $\O(i)\cong\O_Z(i)$. 

We now define the Fourier--Mukai 
functor $\mathbi L$ to be the autoequivalence of $\DC(X)$ 
induced by the kernel ${\cal L}=j_* (\pi^* \O_Z(1)).$
This functor acts on $\DC(X)$ by 
(compare to (\ref{eq:defl}))
\begin{equation}
  \mathbi{L}(\mathsf{B}) = \mathsf{B}\otimes \pi^* \O_Z(1).
\end{equation}
Note that the use of the notation $\mathbi L$ for this functor 
is consistent with the one used in the 
previous section, since in the case when $X$ is a
K3 fibration over $Z \cong \P^1$ we have $\O_X(S)=\pi^* \O_Z(1).$

We also define the exact functor $\mathbi{K}$ induced by the 
kernel ${\cal K}=\Cone(\O_{X \times X} \to \O_\Delta),$ 
with $\O_\Delta :=j_*\O_{X}$ and $\O_{X \times X} \to \O_\Delta$ the
natural restriction map.
We can quote for example Lemma 3.2 
%and Example 3.3
of \cite{ST:braid} to conclude that the action of this functor on
$\DC(X)$ is indeed given by (\ref{eq:mO}).  We make the assumption
that the sheaf $\O_X$ is spherical (as defined in
section \ref{ss:single}). This ensures that the functor $\mathbi{K}$
is an autoequivalence of $\DC(X).$

Finally, we define the exact functor $\mathbi H$ to be the  
so-called fibrewise
Fourier--Mukai transform associated to the Calabi--Yau fibration
$\pi : X \to Z$
(see, for example, \cite{ST:braid},
\cite{BrMa:fibre},  \cite{ACHY:FMug}).
The functor $\mathbi H$ is induced by the
kernel ${\cal H}=\Cone(\O_{X \times_Z X} \to \O_\Delta),$ 
with $\O_{X \times_Z X}$ viewed as a sheaf on $X \times X$
(extension by zero), and $\O_{X \times_Z X} \to \O_\Delta$
the restriction map.

To ensure that the functor $\mathbi H$ is indeed 
an autoequivalence (Fourier--Mukai functor), we assume that 
the sheaf $\O_X$ is EZ--spherical.
In the language of 
\cite{Horj:EZ}, this means that there exists 
a distinguished triangle in $\DC(Z)$ ($Z \cong \P^d$)
of the form\footnote{Lemma 3.12 in \cite{ST:braid}, as well as 
Example 4.2 of \cite{Horj:EZ} provide sufficient conditions 
for (\ref{eq:tr}) to hold.} 

\begin{equation}\label{eq:tr}
\O_Z \to \pi_* \O_X \to \O_Z(-d-1) [-n+d] \to \O_Z [1].
\end{equation}

Note that the sphericity and EZ-sphericity conditions are both
satisfied in the specific example of a K3 fibration over $\P^1$
analyzed in the previous section of this paper.

We now justify the use of the notation $\mathbi H$ to denote the 
fibrewise Fourier--Mukai functor by showing that its action on
$\DC(X)$ is indeed, even in the higher dimensional situation, 
given by the formula (\ref{eq:picon}) of the previous section. 

The fibre product  $X \times_Z X$ fits in the fibre square diagram 

\begin{equation} \label{diag:fq1}
\xymatrix{
X\!\times_Z\! X\ar[d]_{q_2}\ar[r]^(0.6){q_1}&X\ar[d]^{\pi}\\
X\ar[r]^{\pi}&Z\, ,
}
\end{equation}
and let $k : X \times_Z X \hookrightarrow X \times X$ denote the canonical 
embedding. Since $\pi$ is of finite type and flat, we can apply ``cohomology
commutes with the base change'' (Prop. III 9.3 of \cite{Hartshorne:}
or for a more general form Prop. II 5.12 of \cite{Hart:dC}):
\begin{equation}
\pi^*\pi_*\mathsf{B} \cong q_{2*}q_1^*\mathsf{B},
\end{equation}
for some $\mathsf{B}$ in $\DC(X).$
On the other hand, $q_1 = p_1 \circ k,$ and $q_2 = p_2 \circ k,$ so we can write

\begin{equation}
\begin{split}
\pi^*\pi_*\mathsf{B} &\cong
  q_{2*}q_1^*\mathsf{B}\\
  &\cong p_{2*}k_* k^*p_1^*\mathsf{B}\\
  &\cong p_{2*}(k_*\O_{X \times_Z X} \otimes p_1^*\mathsf{B}).
% &\cong \Phi_{k_*\O_{X \times_Z X}}(\mathsf{B}).
\end{split}
\end{equation}
Note that the last line in the previous formula represents the action on $\DC(X)$
of the exact functor $\Phi_{\O_{X \times_Z X}}$
induced as in (\ref{eq:ker}) by the kernel $\O_{X \times_Z X}$ 
(shorthand for $k_*\O_{X \times_Z X}$). 

This shows that 
$\mathbi H (\mathsf{B})=\Cone(\Phi_{\O_{X \times_Z X}}(\mathsf{B})\to \mathsf{B})
%\Phi_{j_*\O_X}(\mathsf{B}))
=\Cone(\pi^*\pi_*\mathsf{B}\to\mathsf{B})$
as desired.

We are now ready to start discussing 
the main goal of this section which
is to prove the following relation between the defined Fourier--Mukai functors:

\begin{equation}\label{eq:pfd}
({\mathbi L}^{-d} \mathbi{K} {\mathbi L}^d)\ldots
(\mathbi{L}^{-1}\mathbi{K}\mathbi{L})\mathbi{K} \cong {\mathbi H}.
\end{equation}

Equivalently, the same formula can be expressed using kernels as

\begin{equation}
({\cal L}^{-d} \star {\cal K} \star {\cal L}^d)\star \ldots \star
({\cal L}^{-1} \star \cal{K} \star \cal{L})\star \cal{K} \cong {\cal H},
\end{equation}
where, for any integer $i,$ ${\cal L}^i = j_* (\pi^* \O_Z (i)).$
Of course, the case $d=1$ ($Z \cong \P^1$) of (\ref{eq:pfd})
is precisely formula (\ref{eq:pf}) of the previous section. 

Before we move on with the technicalities of the
proof, a few remarks are in order. 
Note that the parentheses in the two formulae 
are simply decorative: the composition of functors, as well as the composition 
of kernels are associative (but, of course, not commutative!). For a fixed object 
${\cal G}$ in $ \DC(X\times X),$
the functors ${\cal G} \star -$ and $ -\star{\cal G}$ from
$ \DC(X\times X)$ to $\DC(X\times X)$ are exact functors between
triangulated categories (i.e. they preserve the distinguished triangles). 
Therefore, for any integer $i,$
we can start with the distinguished triangle defining the kernel ${\cal K}$
\begin{equation}
\O_{X \times X} \to \O_\Delta \to {\cal K} \to \O_{X \times X} [1], 
\end{equation}
and apply to it the operations ${\cal L}^{-i} \star - $ and 
$- \star {\cal L}^{i} $ from the left, and right, respectively. 
But 
\begin{equation}
{\cal L}^{-i} \star \O_{X \times X} \star {\cal L}^{i} \cong 
\pi_2^*\O_Z(-i)\otimes\pi_1^*\O_Z(i) \cong
(\pi \times \pi)^* (\O_Z(-i)\boxtimes\O_Z(i)), 
\end{equation}
and 
\begin{equation}
{\cal L}^{-i} \star \O_{\Delta} \star {\cal L}^{i} \cong \O_{\Delta}.
\end{equation}
The notation $\O_Z(-i)\boxtimes\O_Z(i)$ (the exterior tensor product)
will be used 
quite often in what follows and simply designates 
$s_2^* \O_Z(-i)\otimes s_1^*\O_Z(i).$

As a shorthand for later convenience we introduce the following
objects in $\DC(X \times X):$
\begin{equation} \label{ct}
{\cal T}_i:= (\pi \times \pi)^* (\O_Z(-i)\boxtimes\O_Z(i)) \cong 
\pi_2^*\O_Z(-i)\otimes\pi_1^*\O_Z(i)
\end{equation}
and
\begin{equation} 
{\cal C}_i:={\cal L}^{-i} \star {\cal K} \star {\cal L}^{i} \cong
\Cone({\cal T}_i\to\O_\Delta ),
\end{equation}
for $i\in \Z.$ To justify the definition of the kernel ${\cal C}_i,$
we need to explain how to define the morphism ${\cal
T}_i\to\O_\Delta.$ We start with the canonical pairing map on $Z \times
Z$
\begin{equation}
\O_Z(-i)\boxtimes\O_Z(i) \to \O_{\Delta_Z},
\end{equation}
and lift it to $X \times X$ 
\begin{equation}
(\pi \times \pi)^*(\O_Z(-i)\boxtimes\O_Z(i)) \to (\pi \times \pi)^*( \O_{\Delta_Z}).
\end{equation}
We claim that
\begin{equation}\label{eq:fibrpr}
(\pi \times \pi)^*( \O_{\Delta_Z}) \cong \O_{X \times_Z X}. 
\end{equation}
Indeed, for the fibre square 

\begin{equation} 
\xymatrix@C=15mm{
*++{X \times_Z X}\ar@{^{(}->}[d]_k\ar[r]^t&Z\ar[d]^{\Delta_Z}\\
X\!\times \!X\ar[r]^{\pi\times\pi}&Z\times Z\,.
}
\end{equation}
with $t : X \times_Z X \to Z$ the ``diagonal map'' of the fibre square 
(\ref{diag:fq1}), and $\pi \times \pi$ flat, 
we can apply
again ``cohomology commutes with the base change'' to obtain
$(\pi\times\pi)^*(\Delta_Z)_*\O_Z = k_*t^*\O_Z\,$. Since 
$t^*\O_Z = \O_{X \times_Z X},$ the previous formula can be written as
$(\pi\times\pi)^*(\Delta_Z)_*\O_Z =  \O_{X \times_Z X},$ which is exactly 
(\ref{eq:fibrpr}).

The morphism ${\cal T}_i\to\O_\Delta $ is then defined as the composition
\begin{equation}\label{eq:comp}
{\cal T}_i \cong (\pi \times \pi)^* (\O_Z(-i)\boxtimes\O_Z(i)) \to 
(\pi \times \pi)^*( \O_{\Delta_Z}) \cong \O_{X \times_Z X} \to \O_{\Delta}.
\end{equation}
Note that ${\cal C}_0 \cong {\cal K}.$ 

Therefore we have to show that

\begin{equation}\label{eq:aim}
{\cal C}_d \star \ldots \star {\cal C}_1 \star {\cal C}_0 \cong {\cal H}.
\end{equation}
Note that the case $d=0$ is immediate since in that case $Z$ reduces
to a point, and the Fourier--Mukai transforms $\mathbi H$ and $\mathbi K$ coincide.

An important r\^{o}le in what follows will be played by 
Beilinson's resolution of the diagonal in $Z \times Z \cong \P^d \times \P^d$ \cite{Bei:res}

\begin{equation} \label{eq:beil}
0 \to \O_Z(-d) \boxtimes \Omega_Z^d (d) \to \ldots \to
\O_Z(-1) \boxtimes \Omega_Z^1 (1) \to \O_{Z \times Z} \to \O_{\Delta_Z} \to 0,
\end{equation}
where $\Omega_Z^i$ is the sheaf of holomorphic $i$-forms on $\P^d.$ 

For any integer $i,$ $0 \leq i \leq d,$ 
we define the complexes ${\cal S}_i'$ on $Z \times Z$ to be the following 
truncated versions of Beilinson's resolution 
\begin{equation}\label{eq:defs}
0 \to \O_Z(-i) \boxtimes \Omega_Z^i (i) \to \ldots \to
\O_Z(-1) \boxtimes \Omega_Z^1 (1) \to \O_{Z \times Z} \to 0,
\end{equation}
arranged such that the sheaf $\O_{Z \times Z}$ is located at the 0-th position.
Define 
\begin{equation}\label{eq:defss}
{\cal S}_i :=  (\pi \times \pi)^* ({\cal S}_i').
\end{equation}

We claim that there exists a natural map 
\footnote{In fact, by assuming that the sheaf
$\O_X$ is EZ--spherical, it can be shown
that $\Hom_{X \times X}({\cal S}_i, \O_{\Delta})
\cong {\mathbb C}.$ Therefore, the described nonzero morphism from
${\cal S}_i$ to $\O_\Delta$ is essentially unique.}
\begin{equation}
{\cal S}_i \to \O_\Delta,
\end{equation}
that can be defined at the level of complexes, where, as usual, $\O_\Delta$ denotes
the complex on $X \times X$ with the only non-zero component located at the 0-th position.
To see this, we first make the remark that there exists a map 
of complexes ${\cal S}_i' \to \O_{\Delta_Z}.$ Such a map of complexes 
is well defined, since 
the complex ${\cal S}_i'$ is a piece of Beilinson's
resolution. We can now proceed as in (\ref{eq:comp}) and define the desired morphism 
in $\DC(X \times X)$ as the composition
\begin{equation}\label{eq:compp}
{\cal S}_i = (\pi \times \pi)^* ({\cal S}_i') \to 
(\pi \times \pi)^*( \O_{\Delta_Z}) \cong \O_{X \times_Z X} \to \O_{\Delta}.
\end{equation}

The key result to be proved in this section is the following: 

\

{\it 
\underline{Claim}: For any integer $i,$ $0 \leq i \leq d,$ 

\begin{equation}
{\cal C}_i \star \ldots \star {\cal C}_1 \star {\cal C}_0 \cong
\Cone( {\cal S}_i \to \O_\Delta).
\end{equation}
}
Before proving it, let us convince ourselves that 
the claim implies (\ref{eq:aim}). 
The complex ${\cal S}_d'$ is quasi-isomorphic to the sheaf $\O_{\Delta_Z}$ (more
precisely, to the complex on $Z \times Z$ having the sheaf $\O_{\Delta_Z}$ at the 0-th position).
Therefore, the $i=d$ case of the claim states that 
\begin{equation}\label{eq:tri1}
{\cal C}_d \star \ldots \star {\cal C}_1 \star {\cal C}_0 \cong 
\Cone ( (\pi \times \pi)^* (\O_{\Delta_Z}) \to \O_\Delta ).
\end{equation}
Since by (\ref{eq:fibrpr}) $ (\pi \times \pi)^* (\O_{\Delta_Z}) \cong \O_{X \times_Z X},$
we see that the kernel ${\cal C}_d \star \ldots \star {\cal C}_1 \star {\cal C}_0$ is indeed
isomorphic to $\cal H.$

We now proceed with the inductive proof of the claim. The induction is performed 
with respect to $i.$ The case $i=0$
is clear, since ${\cal S}_0'= \O_{Z \times Z}$ and ${\cal S}_0= (\pi\times\pi)^* (\O_{Z \times Z}) = 
\O_{X \times X}.$

To prove the inductive step $i \Rightarrow (i+1),$ we start with 
the natural maps ${\cal S}_i\to\O_\Delta $ and ${\cal T}_{i+1} \to\O_\Delta.$
Applying the
{\em functors} $-\star {\cal S}_i$ and ${\cal T}_{i+1}  \star -$ we get two more maps, and we
can form the commutative square
\begin{equation} 
\begin{split}\label{c2}
\xymatrix{
{\cal T}_{i+1} \star {\cal S}_i \ar[r]\ar[d] &  {\cal S}_i \ar[d]  \\
{\cal T}_{i+1} \ar[r]  & \O_\Delta.
}
\end{split}
\end{equation}

There is a nice result due to Verdier,  guaranteeing that a commuting square 
\begin{equation}\label{c23} 
\begin{split}
\xymatrix{
X'\ar[r]\ar[d] & Y'\ar[d] \\
X'\ar[r]& Y
}
\end{split}
\end{equation}
extends to a ``9--diagram'' of the form
\begin{equation} \label{c34}
\begin{split}
\xymatrix@=6mm{
X'\ar[r]\ar[d] & Y'\ar[r]\ar[d] & Z'\ar[r]\ar[d] &X'\ar[d][1] \\
X\ar[r]\ar[d] & Y\ar[r]\ar[d] & Z\ar[r]\ar[d] &X\ar[d][1] \\
X''\ar[r]\ar[d] & Y''\ar[r]\ar[d] & Z''\ar[r]\ar[d] &X''\ar[d][1] \\
X'\ar[r][1] & Y'\ar[r][1] & Z'\ar[r][1] &X'[2]\, ,\\
}
\end{split}
\end{equation}
where all the rows and columns are  distinguished
triangles, every square commutes, except for the last one (containing the shift operator
$[2]$), which anticommutes (for more details see page 24 in \cite{BBD:perv}). 

Applying Verdier's ``9--diagram'' construction to (\ref{c2}) yields
\begin{equation} \label{c3}
\xymatrix@=6mm{
{\cal T}_{i+1}\star {\cal S}_i \ar[r]\ar[d] & {\cal S}_i  \ar[r]\ar[d]  & {\cal C}_{i+1}\star {\cal S}_i
\ar[r]\ar[d]  
& {\cal T}_{i+1}\star {\cal S}_i [1]\ar[d] \\
{\cal T}_{i+1} \ar[r] \ar[d] & \O_\Delta \ar[r] \ar[d] & {\cal C}_{i+1} \ar[d] \ar[r] & 
{\cal T}_{i+1} [1] \ar[d]  \\
{\cal T}_{i+1}\star ({\cal C}_i \star \ldots \star {\cal C}_0) \ar[d] \ar[r]  & 
{\cal C}_i \star \ldots \star {\cal C}_0 \ar[r] \ar[d] & {\cal C}_{i+1}\star 
({\cal C}_i \star \ldots \star {\cal C}_0)  \ar[r] \ar[d]
& {\cal T}_{i+1}\star ({\cal C}_i \star \ldots \star {\cal C}_0)[1] \ar[d] \\
{\cal T}_{i+1}\star {\cal S}_i[1] \ar[r] & {\cal S}_i[1]  \ar[r]  & {\cal C}_{i+1}\star {\cal S}_i [1] \ar[r] 
& {\cal T}_{i+1}\star {\cal S}_i [2]\,. \\
}
\end{equation}
We are interested
in the term ${\cal C}_{i+1}\star {\cal C}_i \star \ldots \star {\cal C}_0.$ 

To compute it, return for a moment to the commutative diagram (\ref{c34}), and consider 
the ``diagonal'' map  $Y \to Z''.$ The axioms of a triangulated
category guarantee that the morphism $Y \to Z''$ can be included 
in a distinguished triangle of the form 
\begin{equation}
A \to Y \to Z'' \to A[1].
\end{equation}
A crucial piece of the proof of
Verdier's ``9--diagram'' (page 24 in \cite{BBD:perv}) provides another distinguished triangle
that involves $A,$ namely
\begin{equation} \label{c5}
A \to X''\to Y' [1] \to A[1]\,.
\end{equation}

Returning to diagram (\ref{c3}), we obtain the distinguished triangles
\begin{equation}\label{eq:triang1}
{\cal X} \to \O_\Delta \to {\cal C}_{i+1}\star {\cal C}_i \star \ldots \star {\cal C}_0 \to {\cal X}[1]\, , 
\end{equation}
and 
\begin{equation} \label{eq:triang2}
{\cal X}
\to {\cal T}_{i+1}\star {\cal C}_i \star \ldots \star {\cal C}_0 \to 
{\cal S}_i [1] \to 
{\cal X} [1]\, ,
\end{equation}
for some element $\cal X$ in $\DC(X \times X).$

We plan on using the latter triangle to compute the object ${\cal X},$ but for that we need to first 
understand the term 
${\cal T}_{i+1}\star {\cal C}_i \star \ldots \star {\cal C}_0 .$ The leftmost column
of diagram (\ref{c3}) shows that
\begin{equation}
{\cal T}_{i+1}\star {\cal C}_i \star \ldots \star {\cal C}_0 \cong \Cone(
{\cal T}_{i+1}\star {\cal S}_i \to {\cal T}_{i+1}).
\end{equation}
By definition   $ {\cal T}_{i+1}\star {\cal S}_i
={p_{13*}} (p_{23}^*{\cal T}_{i+1} \otimes
p_{12}^* {\cal S}_i)$. After inspecting the definitions of the kernels ${\cal T}_{i+1}$ and ${\cal S}_i,$
it is not hard to see that computing ${\cal T}_{i+1}\star {\cal S}_i$ 
requires the calculation of kernels of the type 
\begin{equation}
\begin{split}
&{p_{13*}} (p_{23}^* (\pi_3^* \O_Z(-i-1) \otimes \pi_2^* \O_Z(i+1))
\otimes p_{12}^* 
(\pi_2^* \O_Z(-j)\otimes \pi_1^*\Omega_Z^j(j))) \cong \\
&\cong (\pi \times \pi)^*  \O_Z(-i-1) \boxtimes \big( \Omega_Z^j(j) \otimes 
%{p_{13*}} p_2^* (\pi^* \O_Z (i+1-j)).
(\Gamma_X)_* (\pi^* \O_Z (i+1-j)) \big),
\end{split}
\end{equation}
with $0 \leq j \leq i,$ and $\Gamma_X : X \to \{ pt \}$ the projection to a point.

But $(\Gamma_X)_* (\pi^* \O_Z (i+1-j)) \cong (\Gamma_Z)_* (\pi_* \O_X \otimes \O_Z (i+1-j)).$ 
Since $\O_X$ is EZ--spherical, the long exact cohomology sequence induced by the 
distinguished triangle (\ref{eq:tr}) implies that
\begin{equation}
(\Gamma_Z)_* (\pi_* \O_X \otimes \O_Z (i+1-j)) \cong \Hom_Z (\O_Z, \O_Z (i+1-j)).
\end{equation}

Summing up our work, we can conclude that
\begin{equation}
\Cone(
{\cal T}_{i+1}\star {\cal S}_i \to {\cal T}_{i+1}) \cong (\pi \times \pi)^* 
(\O_Z(-i-1) \boxtimes {\cal U}_i),
\end{equation}
where ${\cal U}_i$ is the following complex in $\DC(Z)$ 
\begin{equation}
\begin{split}
0 &\to \Omega_Z^i (i) \otimes \Hom_Z (\O_Z, \O_Z (i+1-i) ) \to \ldots \to 
\Omega_Z^j (j) \otimes \Hom_Z (\O_Z, \O_Z (i+1-j) ) \to 
\\
\ldots &\to \Omega_Z^1 (1) \otimes \Hom_Z (\O_Z, \O_Z (i+1-1) ) 
\to \O_Z \otimes \Hom_Z (\O_Z, \O_Z (i+1) ) 
\to \O_Z (i+1) \to 0, 
\end{split}
\end{equation}
arranged such that the sheaf $\O_Z (i+1)$ is located at the 0-th position.

But {\it what} is the complex ${\cal U}_i$ after all? Again the answer 
can be obtained by employing Beilinson's resolution. Consider the following 
two {\it quasi-isomorphic} complexes 
obtained by truncating (\ref{eq:beil}) at the appropriate place:
\begin{equation}
\begin{split}
0 \to \Omega_Z^d (d) \boxtimes \O_Z(-d)
\to \ldots \to \Omega_Z^{i+1} (i+1) &\boxtimes \O_Z(-i-1)
\to 0, \\
0 \to \Omega_Z^i (i) &\boxtimes \O_Z(-i)
\to \ldots \to 
\O_{Z \times Z} \to \O_{\Delta_Z} \to 0,
\end{split}
\end{equation}
arranged such that the sheaf $\O_{\Delta_Z}$ in the second complex is located at the 0-th position
(and, as a result, the sheaf $\Omega_Z^{i+1} (i+1) \boxtimes \O_Z(-i-1)$ is located at the 
$(-i-1)$-th position in the first complex).
Call these complexes ${\cal A}_i$ and ${\cal B}_i,$ respectively. 

On one hand, the well known properties of the cohomology groups 
of the projective space $Z \cong \P^d,$ give that
\begin{equation}
s_{2_*} ({\cal B}_i \otimes  s_1^* \O_Z(i+1)) \cong {\cal U}_i \, .
\end{equation}
On the other hand, the same cohomology properties give that
\begin{equation}
s_{2_*} ({\cal A}_i \otimes  s_1^* \O_Z(i+1)) \cong \Omega_Z^{i+1} (i+1) [i+1].
\end{equation}
But ${\cal B}_i$ and ${\cal A}_i$ are quasi-isomorphic
(i.e. isomorphic in $\DC(Z \times Z)$), hence
\begin{equation}
{\cal U}_i \cong \Omega_Z^{i+1} (i+1) [i+1],
\end{equation}
and 
\begin{equation}
{\cal T}_{i+1}\star {\cal C}_i \star \ldots \star {\cal C}_0 \cong 
\Cone(
{\cal T}_{i+1}\star {\cal S}_i \to {\cal T}_{i+1}) \cong (\pi \times \pi)^* 
(\O_Z(-i-1) \boxtimes \Omega_Z^{i+1} (i+1)) [i+1]\,. 
\end{equation}

The distinguished triangle (\ref{eq:triang2}) and the definition (\ref{eq:defss}) 
of the complexes ${\cal S}_i$ show that in fact our unknown complex $\cal X$
is nothing else but ${\cal S}_{i+1}.$ The proof by induction 
of the main claim of this section 
is then finished by invoking the distinguished triangle (\ref{eq:triang1}).

%%%%%%%%%%%%%%%%%%%%%%%%%%%%%%%%%%%%%%%%%%%%%%%%%%%%%%%%%%%%%%%%%%%

\section{Applications}   \label{s:app}

\subsection{$Z$ is a point}  \label{ss:pt}

It was discussed above that the case of $Z$ being a point amounts to a
single D-brane becoming massless. This is the case originally studied
by Strominger in \cite{Str:con} and yielding monodromy of the form
studied in detail by Seidel and Thomas \cite{ST:braid}.

There are three possibilities:
\begin{enumerate}
  \item $E=X$ in which case we are looking at the primary component of
  the discriminant. The quintic was studied at length in \cite{AD:Dstab}.
  \item $E$ is a complex surface of codimension one in $X$. This could
  arise from the blow-up of an isolated quotient singularity. This was
  studied for example in \cite{DG:fracM}.
  \item $E$ is a rational curve. This is the flop case and was studied
  in \cite{me:point}.
\end{enumerate}

%%%%%%%%%%%%%%%%%%%%%

\subsection{$Z$ is a curve}  \label{ss:cv}

Now we have an infinite number of massless D-branes arising from the
derived category of an algebraic curve. There are two possibilities:
\begin{enumerate}
  \item $E=X$ in which case $X$ is a K3-fibration and
  $Z\cong\P^1$. This is the case we studied in section \ref{s:mon}.
  \item $E$ is a ruled surface arising from blowing up a curve of
  quotient singularities in $X$.
\end{enumerate}
In either case we are essentially looking at nonperturbatively
enhanced gauge symmetry \cite{me:en3g,KMP:enhg}.

Putting $\mathsf{z}=\O_p$ for some point $p\in Z$ we obtain a soliton
$q^*\mathsf{z}=\O_F$ which is the structure sheaf of a single fibre
$F$ of the map $q$. These correspond to the charged vector bosons
responsible for the enhanced gauge symmetry. Clearly these bosons
classically have a moduli space given by $Z$ since $p$ may vary in
$Z$. Upon including quantum effects this leads to a number of massless
hypermultiplets in the theory given by the genus of $Z$
\cite{KMP:enhg,W:MF}.

Putting $\mathsf{z}=\O_Z$ we obtain $q^*\mathsf{z}=\O_X$ which
corresponds to one of the massless monopoles.

In fact we may analyze the complete spectrum of
solitons in Seiberg--Witten theory \cite{SW:I} by using the
``geometric engineering'' approach of \cite{KKL:limit} to ``zoom in''
on the point in the moduli space where the nonabelian gauge symmetry appears.
We will explain exactly what happens  in detail in \cite{AK:Dsu2}.

It is worth speculating that such analysis of Seiberg--Witten theory
may shed some light on the ``local mirror symmetry'' story of papers
such as \cite{KMV:l-mir1}. The massless B-type D-branes associated with the
derived category of $Z$ may be related to the A-type D-brane story of
\cite{KLM:hSW} by some kind of local homological mirror symmetry.

\iffigs
\begin{figure}
  \centerline{\epsfxsize=5cm\epsfbox{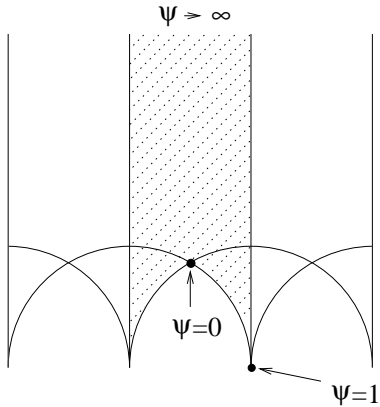}}
  \caption{Moduli Space for Elliptic Curve.}
  \label{f:wl}
\end{figure}
\fi

\subsection{$Z$ is a surface} \label{eq:sf}

There is only one possibility, namely $X=E$ is an elliptic fibration
over $Z$. We will denote this fibration $\pi:X\to S$. Let
$\mathsf{z}\in\DC(S)$ correspond to the skyscraper sheaf of a point
$s\in S$. Then $\pi^*\mathsf{z}\in\DC(X)$ corresponds to the structure
sheaf of an elliptic fibre $e\subset X$ over $s$.

Let us consider the case where the size of $S$ becomes infinite.
According to our rules then, the 2-brane wrapping $e$ should become
massless when we hit the discriminant moving from the large radius
phase to the phase where the elliptic fibres such as $e$ have
collapsed. At first sight this looks peculiar. One does not usually
expect a 2-brane wrapped around a 2-torus to become massless for a
particular radius of the torus!

Actually we will argue that this indeed happens and that it is when
the 2-torus is zero sized that the 2-brane becomes massless. Why
T-duality doesn't interfere with this will become
apparent.\footnote{We are grateful to R.~Plesser for an invaluable
discussion on this point.}

As a specific example let $X$ be given by the following equation in
$\P^4_{\{9,6,1,1,1\}}$:
\begin{equation}
  x_1^2+x_2^3+x_3^{18}+x_4^{18}+x_5^{18}.
\end{equation}
This has a quotient singularity which may be resolved with an
exceptional divisor $\P^2$. One may regard $[x_3,x_4,x_5]$ as
homogeneous coordinates on this $S\cong\P^2$. Fixing a point on $S$
one then has an elliptic fibre $e$ given by a sextic in
$\P^2_{\{3,2,1\}}$. See, for example, \cite{VW:pairs} for more details
about such fibrations.

The moduli space of interest to us regards the area of $e$. This area
is infinite for the \CY\ limit point and shrinks down as we approach
the other limit point. For a more precise statement let us consider
the mirror $\tilde e$ of $e$ given by the following equation
in $\P^2_{\{3,2,1\}}$:
\begin{equation}
  x_1^2+x_2^3+x_3^6+432^{\frac16}\psi x_1x_2x_3.
\end{equation}
Varying the size of $e$ is then mirror to varying the complex
structure of $\tilde e$ by varying $\psi$. Going through the usual
story of solving the Picard--Fuchs equation and using the mirror map
to map back to the $B+iJ$ plane for $e$ we obtain the shaded region in
figure~\ref{f:wl}.

The result is that for the \CY\ limit point we have $\psi\to\infty$
and thus $J\to\infty$ as expected. For the other limit point $\psi=0$
and $J=\sqrt{3}/2$. The discriminant is given by $\psi=1$ and
corresponds to $J=0$, i.e., $e$ has zero area. This is where the
2-branes wrapping $e$ (and all the other D-branes given by
$\pi^*\mathsf{z}$ for $\mathsf{z}\in\DC(S)$) become massless.

So what about T-duality for $e$? Note that the moduli space in
figure~\ref{f:wl} corresponds to {\em two\/} fundamental regions for
the action of $\Sl(2,\Z)$ on the upper-half plane. T-duality should
map the point at $\psi=1$ to the point at $\psi=\infty$. It is
important to remember however that T-duality does not act only upon
$e$ --- one must also shift the string dilaton by an amount related to the
resulting change in the area of the torus. As such, the point at
$\psi=1$ is related to $\psi=\infty$ only with the string coupling
shifted off to infinity, implying again that the D-brane mass is zero.

Therefore if we fix the dilaton to be a finite value we cannot use
T-duality to relate $\psi=1$ to any large radius torus. This explains
why we really do have a massless 2-brane appearing wrapped around a
zero-sized torus.

Note also that if we allow the base $S\cong\P^2$ to have finite size
then the T-duality group ceases to exist anyway as in \cite{AP:T}.

The fact that there is a T-duality relating $\psi=\infty$ to $\psi=1$
shows that their physics must be similar. In particular $\psi=1$ must
be an infinite distance away in the moduli space. Indeed, 
in many respects the spectrum of  stable D-branes at $\psi=1$,
coming from the derived category of $\P^2$, must be similar to the
spectrum one would see upon going to a large radius limit. It would be
interesting to investigate this in more detail and generality.

%%%%%%%%%%%%%%%%%%%%%%

\subsection{The Exoflop}  \label{ss:exof}

Finally let us note that not all the walls of the K\"ahler cone
correspond directly to some subspace $E$ collapsing to $Z$. Even so,
it appears that we can still fit many, if not all examples into the
general EZ language. We illustrate this with an ``exoflop''.

Let $X$ be the degree 12 hypersurface in $\P^4_{\{3,3,3,2,1\}}$. Its
mirror, $Y$, then has defining equation
\begin{equation}
  a_0z_1z_2z_3z_4z_5 + a_1z_1^4 + a_2z_2^4 + a_3z_3^4 + 
  a_4z_4^6 + a_5z_5^{12} + a_6z_4^2z_5^8 + a_7z_4^4z_5^4,
\end{equation}
and we may use the following algebraic coordinates on the moduli space
of complex structures of $Y$:
\begin{equation}
  x=\frac{a_1a_2a_3a_6^2}{a_0^4a_5},\quad
  y=\frac{a_4a_6}{a_7^2},\quad
  z=\frac{a_5a_7}{a_6^2}.
\end{equation}

We have chosen coordinates so that the K\"ahler cone of $X$ appears
naturally as a positive octant in the secondary fan. In particular
this means that the 3 rational curves in the moduli space connecting
the large radius limit to each of the three neighbouring phase limit
points are given by setting $x=y=0$, $x=z=0$ or $y=z=0$ respectively.

The discriminant has two components given by
\begin{multline}
\Delta_0= -1-64 x+768 x z+32768 x^2 z-196608 x^2 z^2- 
4194304 x^3 z^2+
294912 x^2 yz^2 \\
+16777216 x^3( y z^2+z^3)-75497472 x^3 yz^3+113246208 x^3 y^2 z^4
\end{multline}
and
\begin{equation}
\Delta_1=-1+4y+4z-18yz+27y^2z^2.
\end{equation}

The phase transition of interest occurs for the rational curve $C$ for
which $y=z=0$. For small $x$ we are in the \CY\ phase. For large $x$
the \CY\ $X$ undergoes an ``exoflop'' \cite{AGM:II}. That is $X$
becomes reducible with one component consisting of a threefold with a
singularity. The other component consists of a fibration of a Landau--Ginzburg
orbifold theory over a $\P^1$. These components intersect at a point
which is the singularity in the threefold component.

This then is not an EZ transformation. Note however that the
discriminant $\Delta_0$ intersects $C$ transversely at
$(x,y,z)=(-\ff1{64},0,0)$ and so the monodromy within $C$ is exactly
given by monodromy around the primary component of the discriminant.
Therefore exactly one D-brane becomes massless at the transition point
--- the D6-brane wrapping $X$. This exoflop transition is equivalent,
as far as monodromy is concerned, to an EZ transformation with $X=E$
and $Z$ given by a point.

This is not entirely surprising given the following. Classically one
would describe the exoflop wall of the K\"ahler cone as a wall where 
$\int_X J^3=0$. Thus the classical volume of $X$ is going to zero,
even though the volume of some surfaces and curves within $X$, as
measured by the K\"ahler form, do not vanish. Since the volume of $X$
vanishes, one should expect the D6-brane to have vanishing mass. The
fact that no other D-branes become massless is not obvious.

It would be interesting to show that all phase transitions give rise
to monodromies that can be associated with EZ transformations.

%%%%%%%%%%%%%%%%%%%%%%%%%%%%%%%%%%%%%%%%%%%%%%%%%%%%%%%%%%%%%%%%%%%

\section*{Acknowledgments}

It is a pleasure to thank J.~Distler, D.~Morrison and R.~Plesser
for useful conversations.  P.S.A.\ is supported in part by NSF grant
DMS-0074072 and by a research fellowship from the Alfred P.~Sloan
Foundation. R.L.K. was partly supported by NSF grants
DMS-9983320 and DMS-0074072.

%\bibliographystyle{my-phys}
%\bibliography{string}

\end{document}